\documentclass[
    reprint,
    superscriptaddress,
    amsmath,
    amssymb,
    aps
]{revtex4-2}

\usepackage{graphicx}
\usepackage[hidelinks]{hyperref}

\begin{document}

\title{Cryogenic Optoelectronic Characterization of InGaAsP LEDs for Low-Energy Detector Calibration}

\author{M.~Ghaith}
\email{Corresponding author: muad.ghaith@zu.ac.ae}
\altaffiliation{Current address: College of Natural and Health Sciences, Zayed University, Dubai, 19282, U.A.E.}
\affiliation{\mbox{Department of Physics, Engineering Physics and Astronomy, Queen's University, Kingston, ON K7L 3N6, Canada}}
\affiliation{\mbox{College of Natural and Health Sciences, Zayed University, Dubai 19282, United Arab Emirates}}

\author{W.~Rau}
\affiliation{\mbox{Department of Physics, Engineering Physics and Astronomy, Queen's University, Kingston, ON K7L 3N6, Canada}}
\affiliation{TRIUMF, Vancouver, BC V6T 2A3, Canada}

\author{P.C.F.~Di~Stefano}
\affiliation{\mbox{Department of Physics, Engineering Physics and Astronomy, Queen's University, Kingston, ON K7L 3N6, Canada}}

\author{M.~Peterson-Galema}
\affiliation{\mbox{Department of Physics, Engineering Physics and Astronomy, Queen's University, Kingston, ON K7L 3N6, Canada}}

\author{E.~Fascione}
\affiliation{\mbox{Department of Physics, Engineering Physics and Astronomy, Queen's University, Kingston, ON K7L 3N6, Canada}}
\affiliation{TRIUMF, Vancouver, BC V6T 2A3, Canada}

\author{R.~Germond}
\affiliation{\mbox{Department of Physics, Engineering Physics and Astronomy, Queen's University, Kingston, ON K7L 3N6, Canada}}
\affiliation{TRIUMF, Vancouver, BC V6T 2A3, Canada}

\author{R.~Underwood}
\affiliation{\mbox{Department of Physics, Engineering Physics and Astronomy, Queen's University, Kingston, ON K7L 3N6, Canada}}
\affiliation{TRIUMF, Vancouver, BC V6T 2A3, Canada}

\date{\today}
\begin{abstract}
We measured the emission spectra and current–voltage (IV) characteristics of commercial InGaAsP LEDs in a cryostat over a temperature range from 7~K to 300~K. At a constant operating current of 3.3~mA, the peak emission wavelength of the 1200~nm (1650~nm) LED decreased from $1202.3 \pm 0.5$~nm ($1665.5 \pm 0.6$~nm) at 300~K to $1143.1 \pm 0.3$~nm ($1539.5 \pm 2.2$~nm) at 7~K, demonstrating a strong temperature dependence. The cut-in voltage increased systematically with decreasing temperature, in agreement with conventional diode theory. To assess junction self-heating effects, we conducted pulsed IV measurements at different duty cycles. The extracted apparent series resistance showed a clear duty-cycle dependence, whereas the cut-in voltage remained unchanged within experimental uncertainty, indicating that self-heating primarily influenced the IV slope rather than the cut-in voltage. These results quantitatively characterize the thermal behavior of both the spectral peak and key electrical parameters, supporting the application of InGaAsP LEDs as cryogenic calibration light sources.


\begin{description}
\item[keywords]
Cryogenic optoelectronics, Optical characterization, Temperature-dependent emission, Detector calibration

\end{description}
\end{abstract}
\maketitle


\section{Introduction}
\label{sec:introduction}

Rare-event searches targeting low-mass dark matter and coherent elastic \mbox{neutrino--nucleus} scattering (CE$\nu$NS) require cryogenic detectors with a well-characterized response at low energy. Light-emitting diodes (LEDs) provide a well-controlled and reproducible method for calibrating the low-energy response of cryogenic detectors at the few-eV scale, while near-infrared LEDs can provide individual photons with energies below 1 eV \cite{DelCastello2024, Ghaith2020, XENON2026LightDM}.

LEDs exhibit a significant shift of their emission wavelength away from the nominal value when operated far below room temperature; ref.~\cite{5lnp-6mng} reports a measured photon energy of $2.05 \pm 0.02$~eV from an LED with a nominal emission wavelength of 630~nm, corresponding to 1.97~eV, when operated while immersed in liquid helium. Ignoring
the shift in emission energy due to the LED operating
temperature introduces a systematic bias if LEDs
operated at cryogenic temperatures are used for detector calibration.

Accurate energy-scale measurements are also crucial for understanding details of the signal formation process in the detector. In semiconductors, charge carriers can measure the energy of a particle interaction, either by direct charge measurements or by exploiting the Neganov-Trofimov-Luke (NTL) effect, which converts electric potential energy into phonons when a voltage is applied to collect the charge carriers. The NTL effect is, e.g.,\ central for the High Voltage eV-resolution (HVeV) detectors developed by SuperCDMS (see e.g.\ \cite{PhysRevD.111.012006}), where it enables single electron-hole pair resolution. However, charge transport, and thus signal formation in both direct charge measurement and phonon-mediated charge measurement, is affected by charge trapping, impact ionization, and charge-carrier recombination. Thanks to an accurate knowledge of the photon emission energy of an LED operated at low temperature, the authors of \cite{Kennard:2026xlx} were able to identify the presence of a signal loss process in the phonon-mediated signal formation. By rigorously characterizing the temperature-dependent shift in LED emission wavelength, we can minimize systematic biases in low-energy detector response modeling. 

In this work, we present a unified optical-electrical characterization of commercial InGaAsP LEDs with nominal peak emissions at 1200~nm and 1650~nm, operated in a cryostat at temperatures from 7~K to 300~K, quantifying the temperature dependence of the emission spectrum, examining current-dependent peak shifts at low temperature, and measuring IV characteristics across the same temperature range to extract the cut-in voltage and apparent series resistance. Together, these results provide device-level information important for using these (or similar) LEDs as controlled cryogenic photon sources for low-energy calibration.


\section{Background}
\label{sec:background}

LEDs are pn-junction devices in which forward bias injects electrons and holes into an active region, where a fraction of recombination events are radiative \cite{neamen2012_semiconductor}. The emitted photon energy is primarily determined by the semiconductor bandgap energy, $E_g$; accordingly, the peak emission wavelength is commonly approximated by

\begin{equation}
\lambda \approx \frac{hc}{E_g},
\end{equation}

\noindent where $h$ is Planck’s constant, $c$ is the speed of light, and $\lambda$ is the emission wavelength \cite{kittel2018introduction,neamen2012_semiconductor}. Because the bandgap is temperature dependent, the emission peak correspondingly shifts with temperature \cite{odonnell1991_bandgaps}. A widely used empirical description of this dependence is Varshni’s relation \cite{varshni1967_energygap}

\begin{equation}\label{eq:Varnish}
E_g(T) = E_0 - \frac{\alpha T^2}{T + \beta},
\end{equation}

\noindent where $E_0$, $\alpha$, and $\beta$ are material-specific parameters. For InGaAsP, these parameters are tabulated in ref.~\cite{LI2000215}.

In actually realized LEDs, however, the measured spectrum may deviate from a single symmetric peak \cite{reifegerste2008_ledspectra} due to band filling, carrier heating, and compositional or structural inhomogeneity within the active region \cite{wu2012_ingan_el}. These deviations motivate direct spectral measurements and a consistent peak-extraction procedure across the investigated temperature range.

Complementing the optical analysis, the IV characteristics provide an electrical probe of junction behavior. Under forward bias, a diode transitions rapidly from weak conduction to substantial current near the so-called turn-on (cut-in) voltage \cite{neamen2012_semiconductor}. Although the cut-in voltage is not a fundamental material constant, it serves as a practical experimental parameter that reflects temperature-dependent changes in the effective junction barrier and band structure \cite{neamen2012_semiconductor,fulmek2017_inganfwd,akimoto1998_ledlowT}. At higher forward currents, the approximately linear region of the IV curve is commonly used to extract an apparent series resistance, which encompasses contributions from the bulk semiconductor, contacts, wiring, and other non-ideal effects \cite{fulmek2017_inganfwd}.

The remainder of this manuscript is organized as follows: section \ref{sec:methodology} describes the experimental setup and measurement procedures; section \ref{sec:results} presents the results and discussion; and section \ref{sec:conclusion} summarizes the main conclusions.


\section{Methodology}
\label{sec:methodology}
\subsection{Experimental setup}

\begin{figure}[t]
    \centering
\includegraphics[width=\linewidth, trim=5mm 1mm 5mm 5mm, clip]
{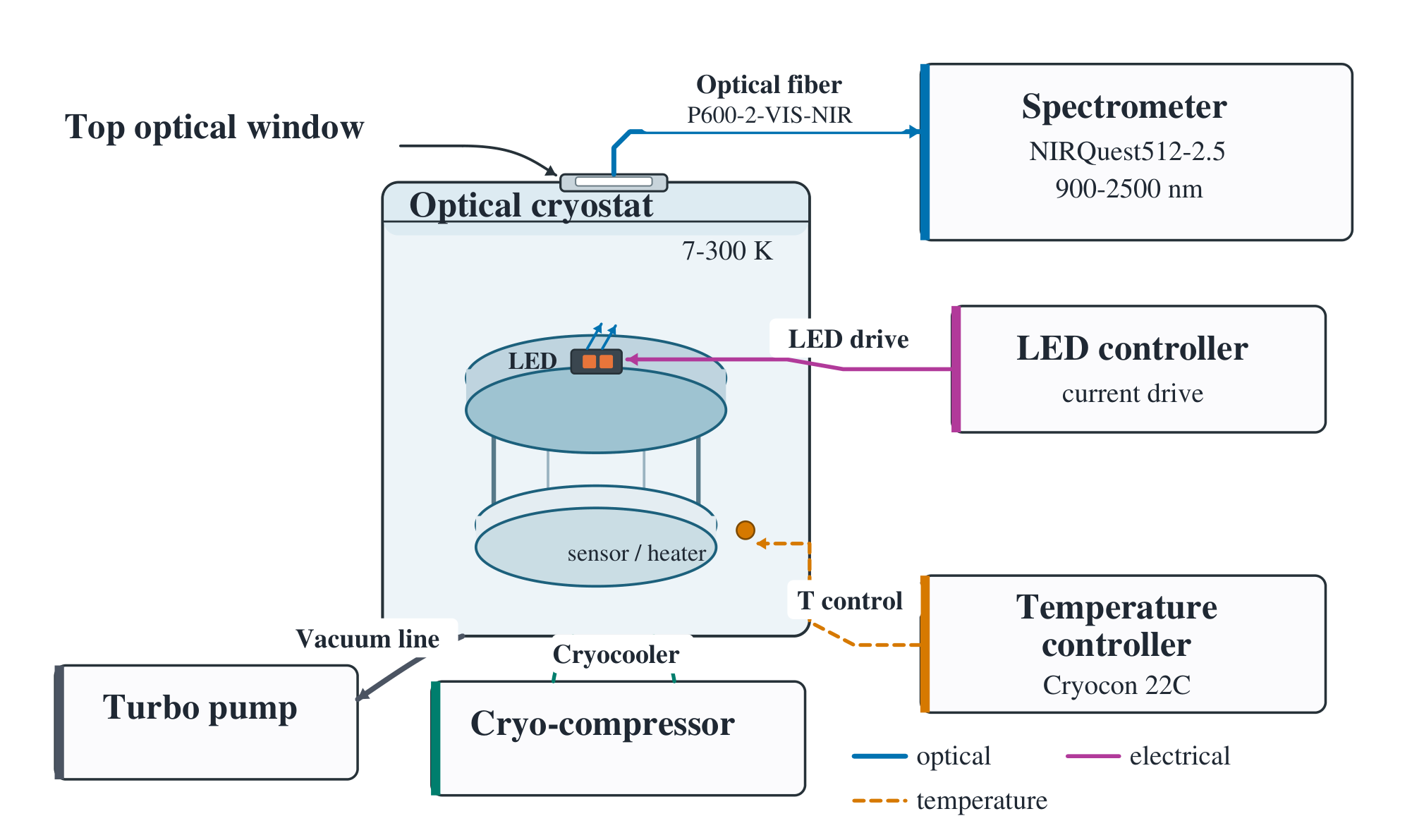}
    \caption{Schematic of the experimental setup showing two LEDs facing the cryostat's top window. The optical fiber is attached to the top window of the cryostat. The LED controller and temperature controller are connected to the cryostat via SMA connectors.}
    \label{fig:schematic}
\end{figure}

We have characterized two commercial InGaAsP LEDs with nominal peak emission wavelengths of 1200~nm and 1650~nm (Marubeni Co.~\cite{SMC1200datasheet}) over the temperature range from 7~K to 300~K. We performed measurements in a closed-loop ColdEdge cryostat cooled by a Sumitomo HC-4A helium compressor to a base temperature of approximately 7~K. Prior to cooling, a Pfeiffer rotary vacuum pump evacuated the sample chamber to pressures below $10^{-3}$~mbar to suppress convective heat transfer. The cold-stage temperature was regulated using a Cryocon 22C controller. A schematic diagram for the setup is shown in Fig.~\ref{fig:schematic}.

\begin{figure}[t]
	\centering
	\includegraphics[width=0.88\columnwidth]{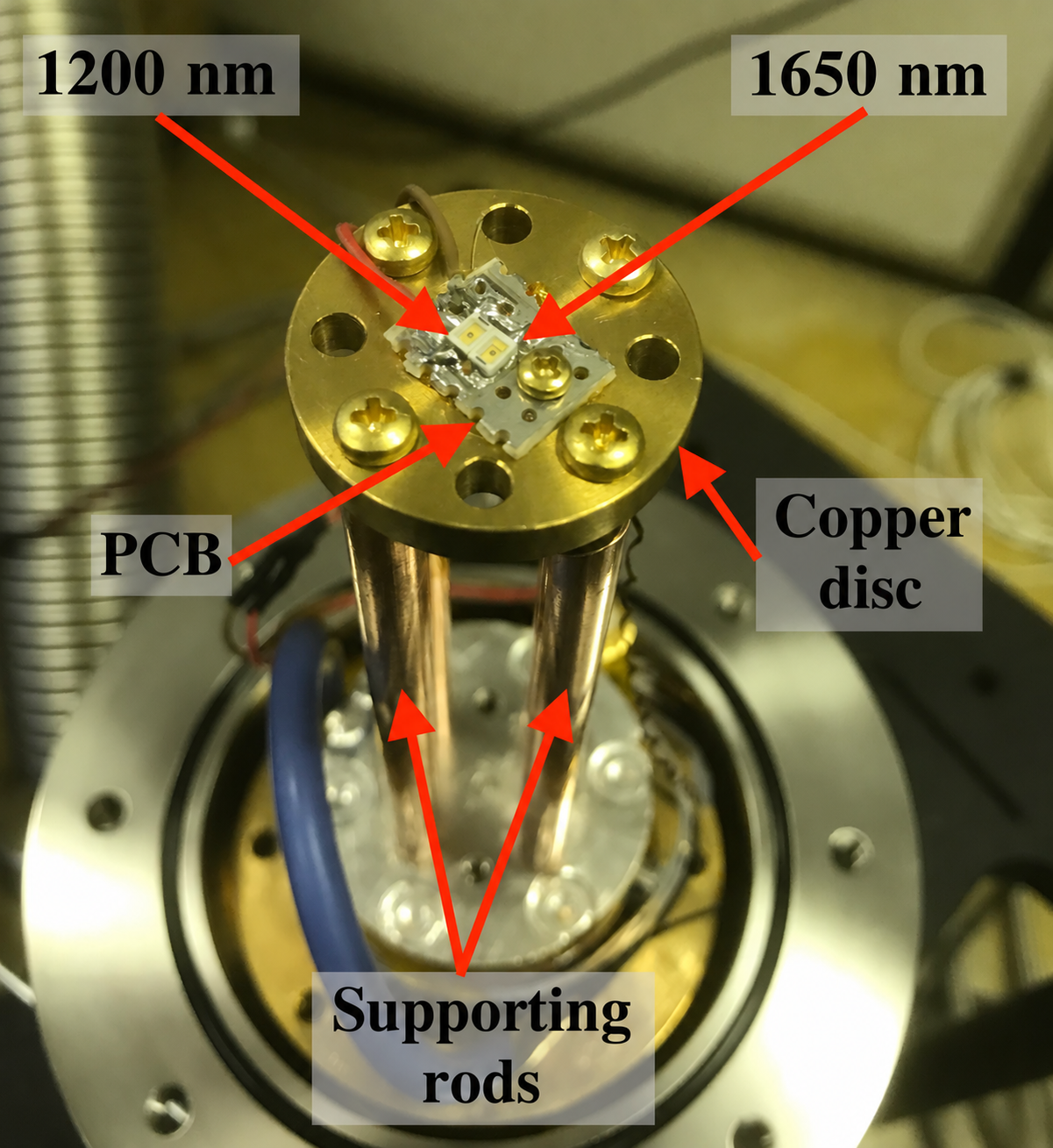}
    \vspace{0.2cm} 
    \caption{Four supporting copper rods attached to the cryostat cold head elevate the LED assembly toward the top window for better photon collection through the optical fiber, which is mounted outside the window.}
	\label{figs:setup1}
\end{figure}

\begin{figure}[t]
	\centering
	\includegraphics[width=0.88\columnwidth]{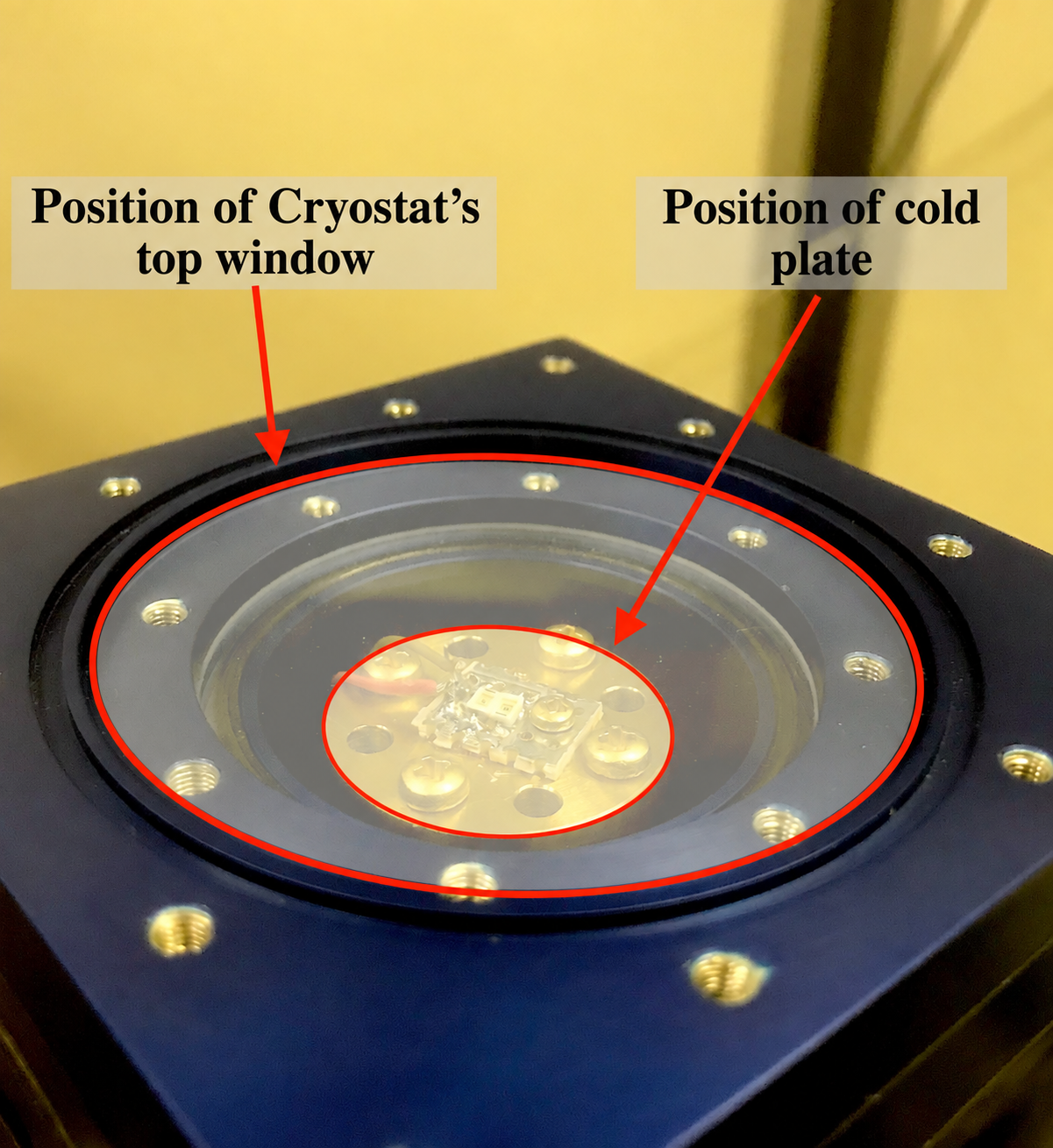}
    \caption{Top view of the closed cryostat before adding the window cover with the optical fiber,  showing the LEDs mounted on a PCB and secured to the cold plate beneath where the optical fiber will be positioned.}
	\label{figs:setup2}
\end{figure}

The LEDs, with outer dimensions of 3.5~mm$\times$2.7~mm, were mounted on ceramic carriers and encapsulated in epoxy resin. We soldered the two LED packages onto a $1\times 1.5~\mathrm{cm}^2$ PCB, which we fixed to a gold-plated copper disk thermally anchored to the cryostat cold head (Fig.~\ref{figs:setup1} and \ref{figs:setup2}). 

The disk was mounted on four 10~cm-long copper rods, positioning the LED assembly within a few millimeters of the infrared-transparent Suprasil window to enhance photon collection by the optical fiber mounted outside the cryostat while ensuring good thermal contact with the cold plate.

\begin{figure}[t]
	\centering
	\includegraphics[width=0.88\columnwidth]{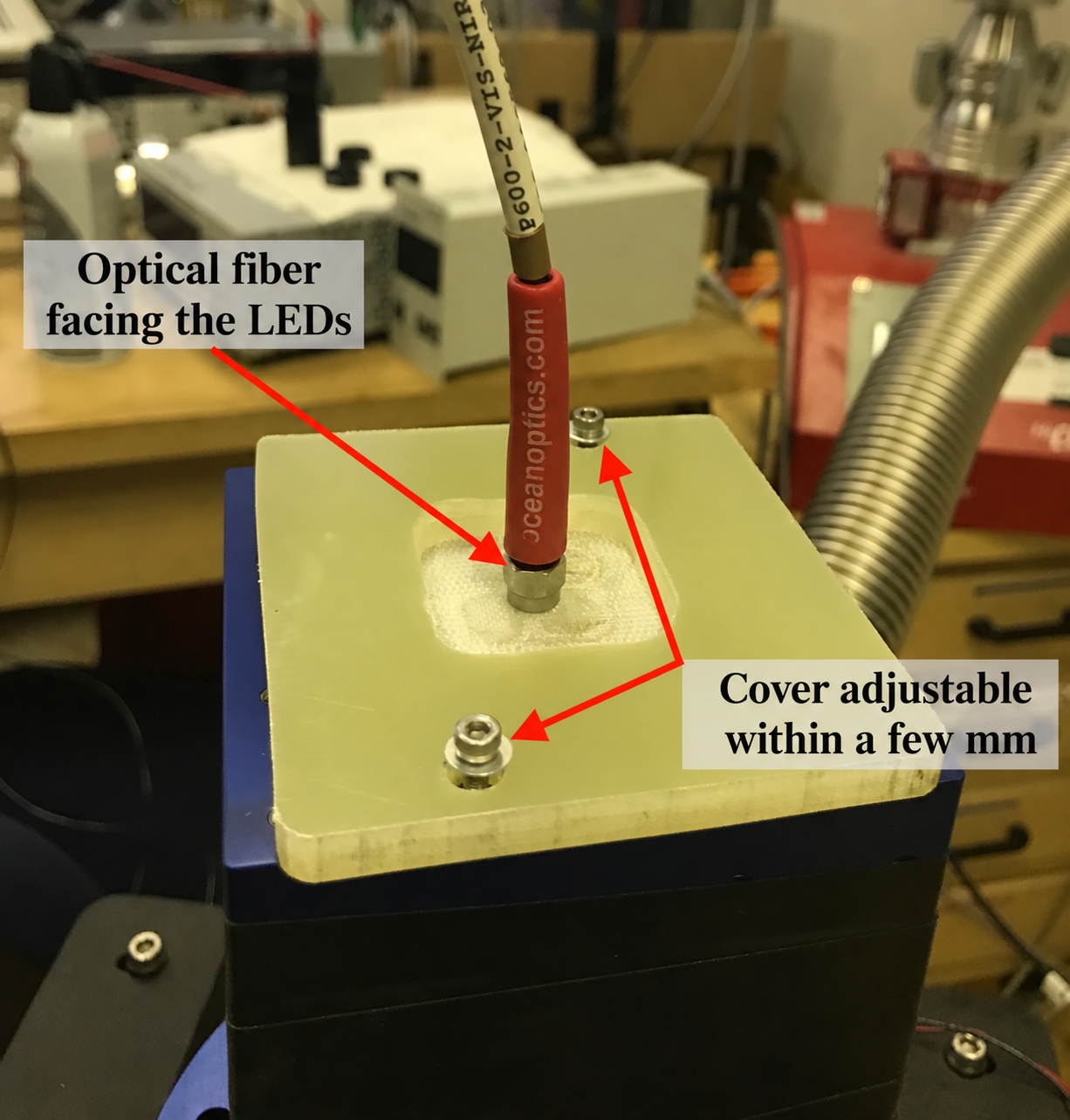}
	\caption{An \(x\)--\(y\)-adjustable mounting plate was used to position the tip of the optical fiber a few millimeters above the LED under investigation, with the fiber directly facing the LED.}
	\label{figs:setup3}
\end{figure}

A thermometer was mounted between one copper spacer and the copper disk to monitor device temperature. Optical emission spectra were recorded using an Ocean Optics NIRQuest512-2.5 spectrometer (900--2500~nm) coupled via a P-600-2-VIS-NIR fiber, routed from the cryostat window, and held in place by a mounting plate attached outside the cryostat that could be aligned to face either of the LEDs (Fig.~\ref{figs:setup3}).

\subsection{Measurement procedure}
Before each measurement cycle, we evacuated and cooled the cryostat from 300~K to its base temperature of approximately 7~K, following the procedure described above. When the temperature reached about 70~K, we closed the vacuum valve and switched off the pump. During warm-up above roughly 100~K, the pump was turned back on to maintain a good vacuum. The temperature was stepped between 7~K and 300~K using the built-in heater. At each step, we allowed the system to equilibrate for typically 15~min (longer for larger temperature changes) until the thermometer reading was stable within 0.04~K.

At each temperature, the IV characteristics were measured by sweeping the applied voltage and recording the corresponding current using the Arduino-based circuit described in Ref.~\cite{Ghaith2020}, while the corresponding LED emission spectra were recorded simultaneously. At selected temperatures, we performed additional spectral measurements under pulsed-bias conditions, systematically varying the pulse width, amplitude, and repetition frequency. Infrared background contributions were subsequently subtracted using OceanView\texttrademark{} software to obtain the corrected emission spectra.


\section{Results and Discussion}
\label{sec:results}

\subsection{IV curve results}

\begin{figure}[t]
	\centering
	\includegraphics[width=\columnwidth]{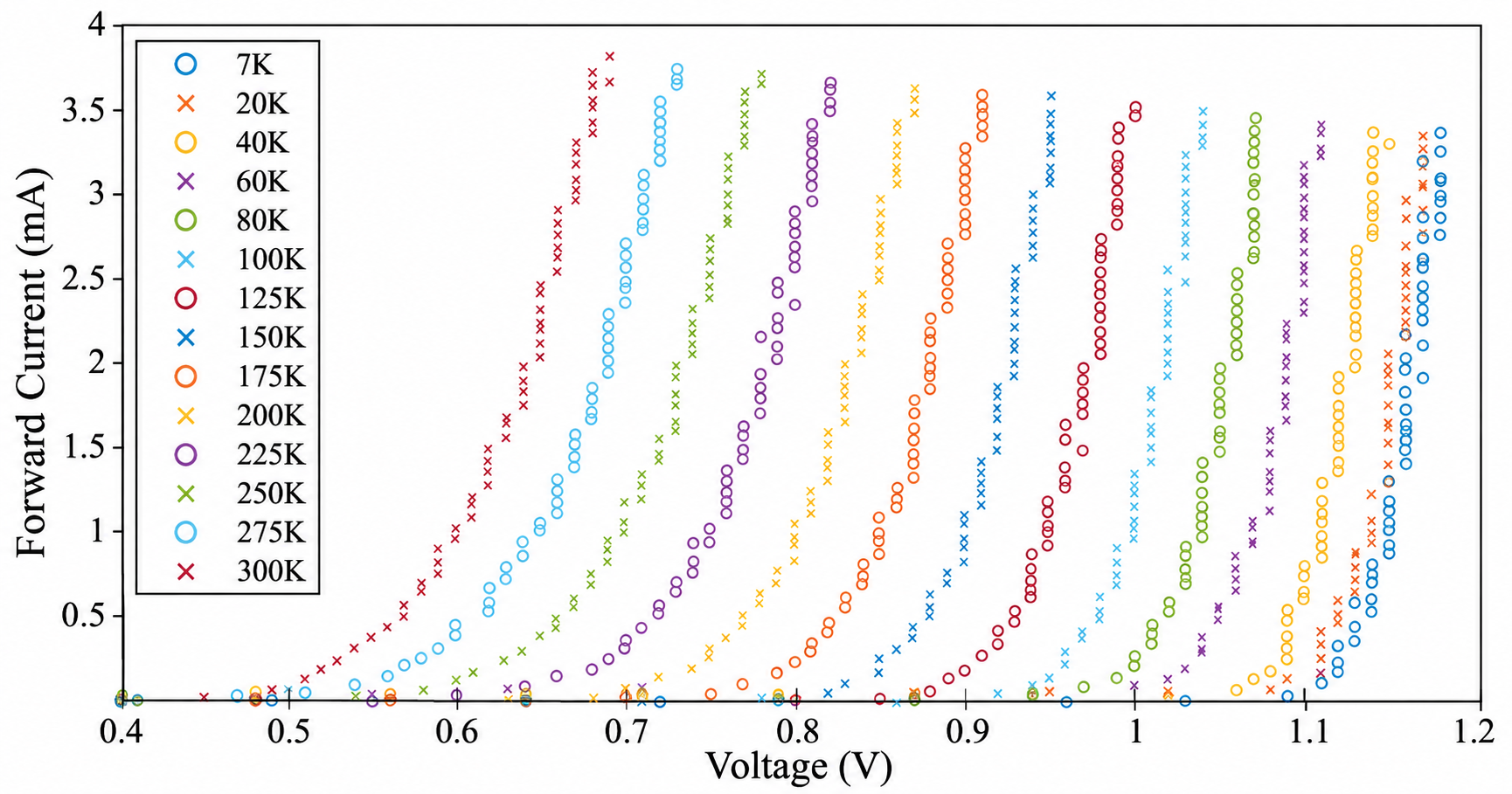} 
    \caption{IV characteristics of the 1650-nm LED measured at temperatures from 7~K to 300~K. A systematic shift of the turn-on voltage and a change in slope with temperature are obvious.}
	\label{fig:iv_all}
\end{figure}

The IV characteristics of the LED in forward-bias mode were measured at 14 temperatures ranging from 7~K to 300~K. For these measurements, short (100~$\mu$s) pulses of varying voltage were generated at a frequency of 1~kHz, and the resulting current was measured for each pulse. At any given temperature, the current first rises slowly with voltage and then enters a roughly linear regime. Fig.~\ref{fig:iv_all} shows the IV curves for the 1650~nm LED as an example. The onset and slope of the IV curves change significantly with temperature.

The slope in the linear region of the IV curve ($\mathrm{d}I/\mathrm{d}V$) corresponds to the differential conductance of the LED. The inverse of that slope ($\mathrm{d}V/\mathrm{d}I$) is the apparent series resistance. To extract the slope, we fitted a straight line to the data and excluded the non-linear region by omitting data points corresponding to currents below 0.75~mA. The cut-in voltage, $V_c$, is defined as the $x$-intercept of the respective fitting lines. 

\begin{figure}[t]
	\centering
	\includegraphics[width=\columnwidth]{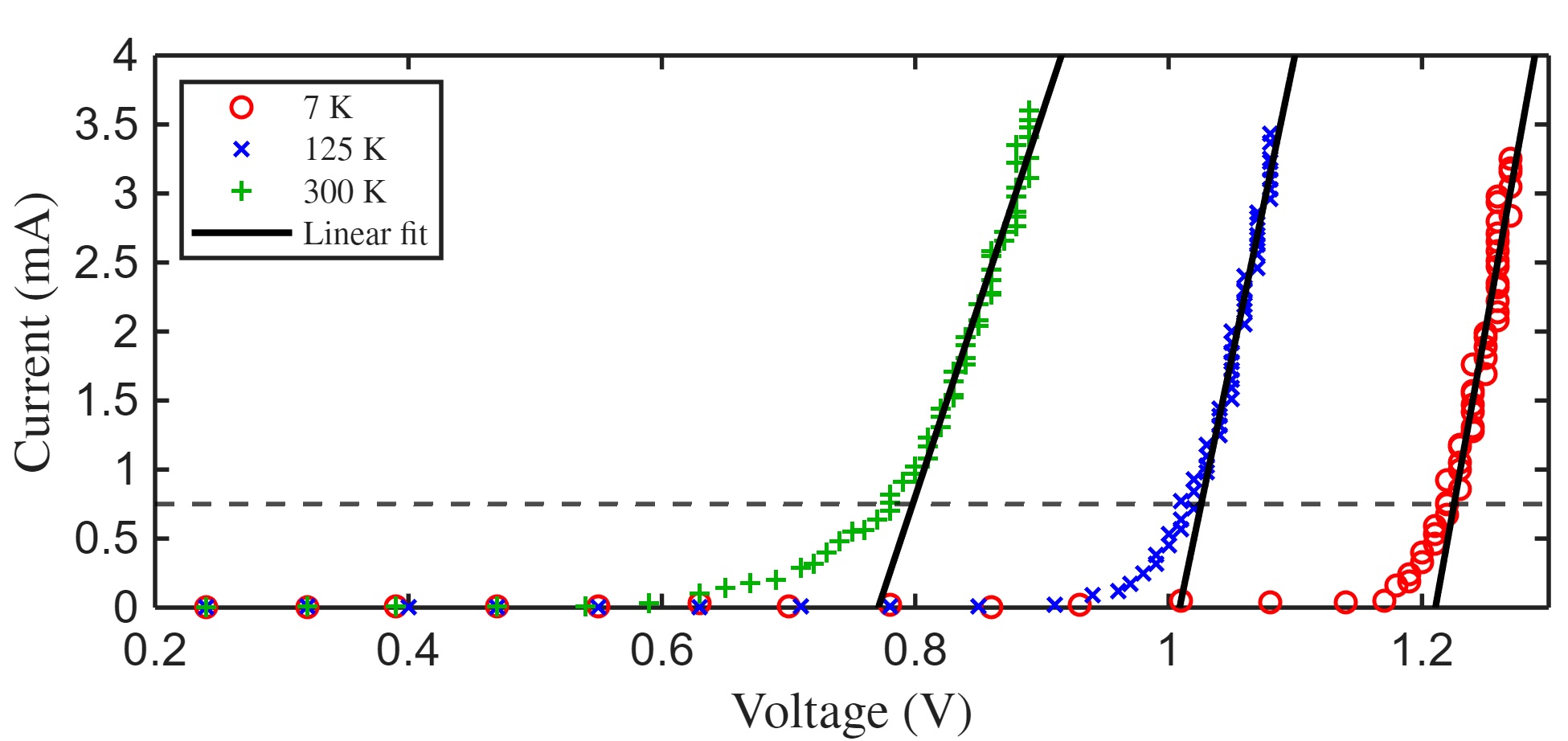} 

    \caption{Representative forward-bias IV curves at 7~K, 125~K, and 300~K (1200-nm LED) illustrating the method used to extract the apparent series resistance and cut-in voltage from the approximately linear high-current region (inverse slope and $x$-axis intercept, respectively, of the fit lines shown in black). The dashed line indicates the current threshold above which data points were used for the fits.}
	\label{fig:iv_fits}
\end{figure}

Representative fits to data from the 1200-nm LED at 7~K (red), 125~K (blue), and 300~K (green) are shown in Fig.~\ref{fig:iv_fits}.

\begin{figure}[t]
	\centering
	\includegraphics[width=\columnwidth]{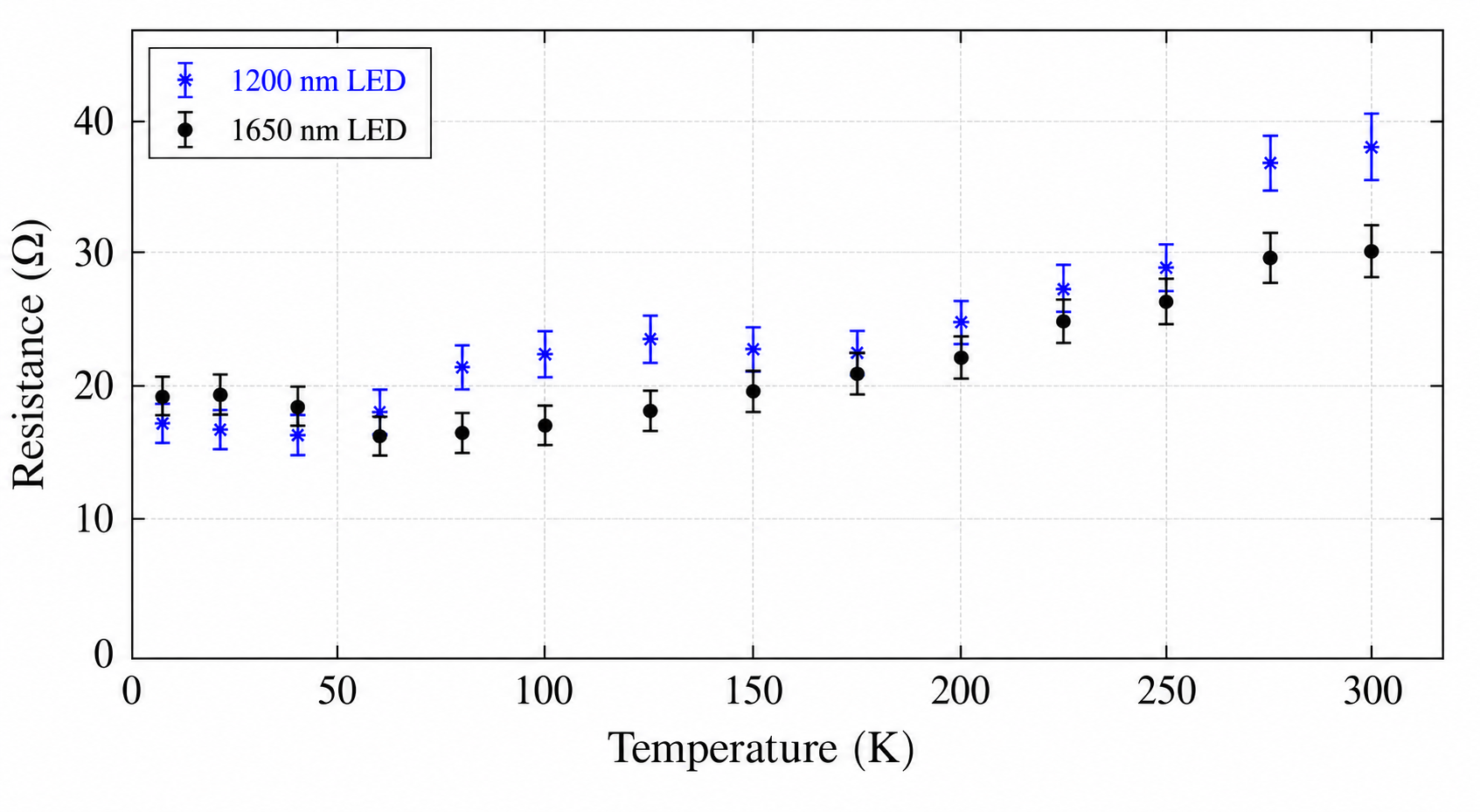}

    \caption{Temperature dependence of the apparent series resistance extracted from the linear high-current region of the IV curves of the forward-biased LEDs.}
	\label{fig:R_vs_T}
\end{figure}

\begin{figure}[t]
	\centering
	\includegraphics[width=\columnwidth]{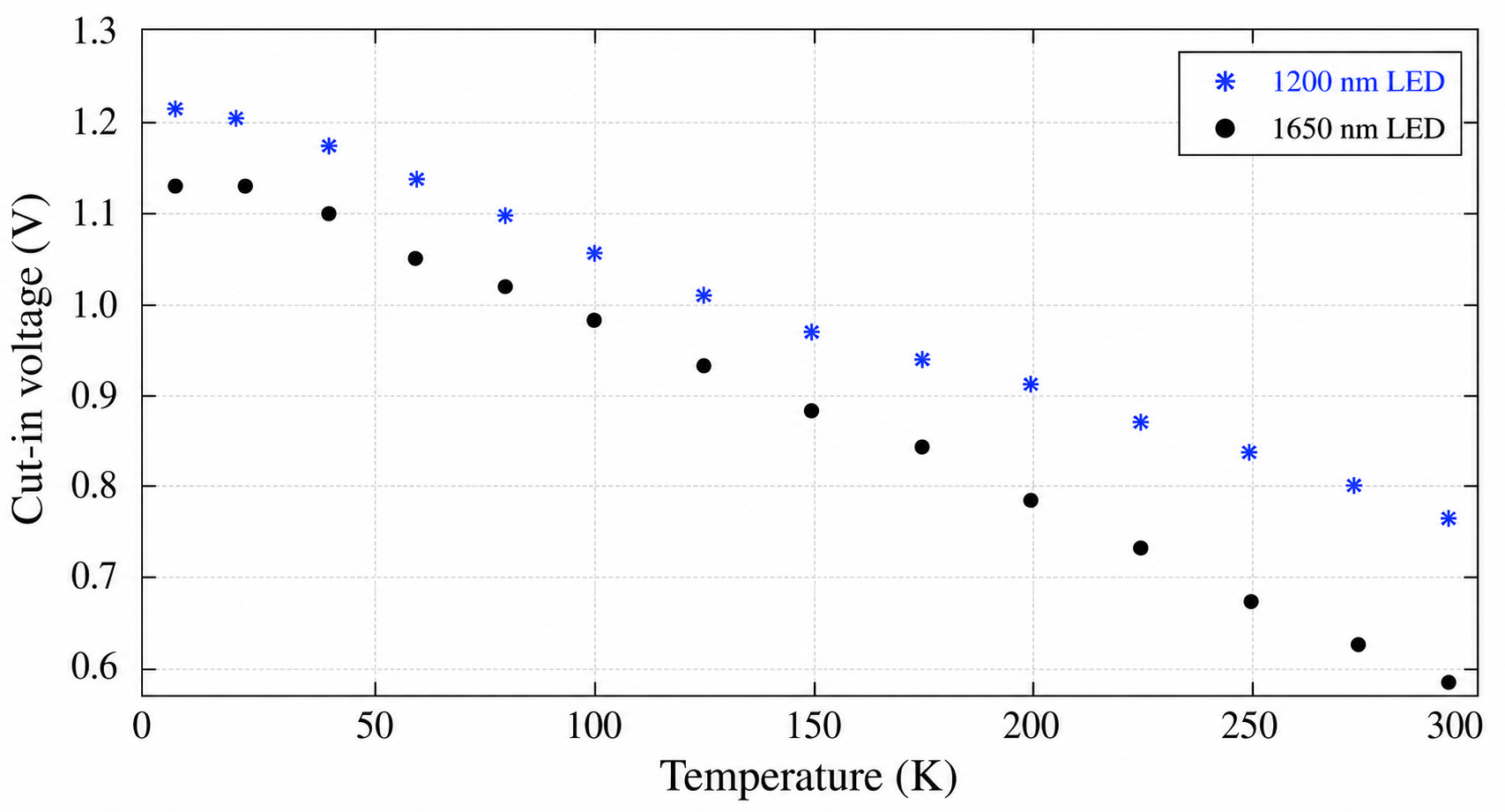}

    \caption{Temperature dependence of the extracted LED cut-in voltage, $V_c$, from 7~K to 300~K, showing a monotonic decrease with increasing temperature. Error bars are smaller than the symbol size and are therefore not visible.}
	\label{fig:vc_vs_T}
\end{figure}

As the temperature decreases, the cut-in voltage increases systematically, from approximately 0.77~V (0.58~V) at 300~K to approximately 1.21~V (1.13~V) at 7~K for the 1200-nm (1650-nm) LED. This shift is consistent with the increase in semiconductor band-gap energy at lower temperatures \cite{band, band2}. The non-linear onset region of the IV curves also steepens as temperature decreases, reflecting reduced thermal broadening and a corresponding decrease in the ideality factor \cite{PhysRevB.95.035211}. At higher temperatures, the broader transition and lower turn-on voltage indicate enhanced thermally activated carrier transport and recombination. These trends are consistent with the expected temperature dependence of carrier injection and recombination processes in direct-band-gap semiconductor LEDs \cite{Ryu:20, Rashidi2019}. Figures~\ref{fig:R_vs_T} and \ref{fig:vc_vs_T}  show the temperature dependence of the apparent series resistance and the cut-in voltage, respectively, for both LEDs. Data are also reported in Table~\ref{tab:R_Vc}.

\begin{table}[t]
\caption{Parameters extracted from the measured IV curves (cut-in voltage $V_c$ and apparent resistance R) as a function of temperature for both LEDs while being operated in continuous mode. Values in parentheses indicate uncertainties in the last digits.}
\label{tab:R_Vc}

\centering
\renewcommand{\arraystretch}{1.1}
\begin{tabular}{ccccc}
\hline
\textbf{T (K)} & \multicolumn{2}{c}{\textbf{1200~nm LED}} & \multicolumn{2}{c}{\textbf{1650~nm LED}} \\
 & $V_c$ (V) & $R$ ($\Omega$) & $V_c$ (V) & $R$ ($\Omega$) \\
\hline
7   & 1.212(4) & 19(2) & 1.128(3) & 17(2) \\
20  & 1.203(5) & 19(2) & 1.125(4) & 14(2) \\
40  & 1.174(5) & 18(2) & 1.095(3) & 15(1) \\
60  & 1.137(5) & 18(2) & 1.057(4) & 16(2) \\
80  & 1.093(4) & 22(2) & 1.022(3) & 16(2) \\
100 & 1.052(4) & 23(2) & 0.981(3) & 17(1) \\
125 & 1.006(4) & 24(2) & 0.933(4) & 19(2) \\
150 & 0.966(4) & 23(2) & 0.883(3) & 20(2) \\
175 & 0.936(4) & 22(2) & 0.832(4) & 23(2) \\
200 & 0.911(4) & 53(2) & 0.782(4) & 25(2) \\
225 & 0.867(5) & 27(2) & 0.728(4) & 26(2) \\
250 & 0.834(5) & 65(2) & 0.674(4) & 29(2) \\
275 & 0.796(5) & 37(2) & 0.624(4) & 29(2) \\
300 & 0.766(6) & 38(3) & 0.578(5) & 30(2) \\
\hline
\end{tabular}
\end{table}

\begin{figure}[t]
	\centering
	\includegraphics[width=\columnwidth]{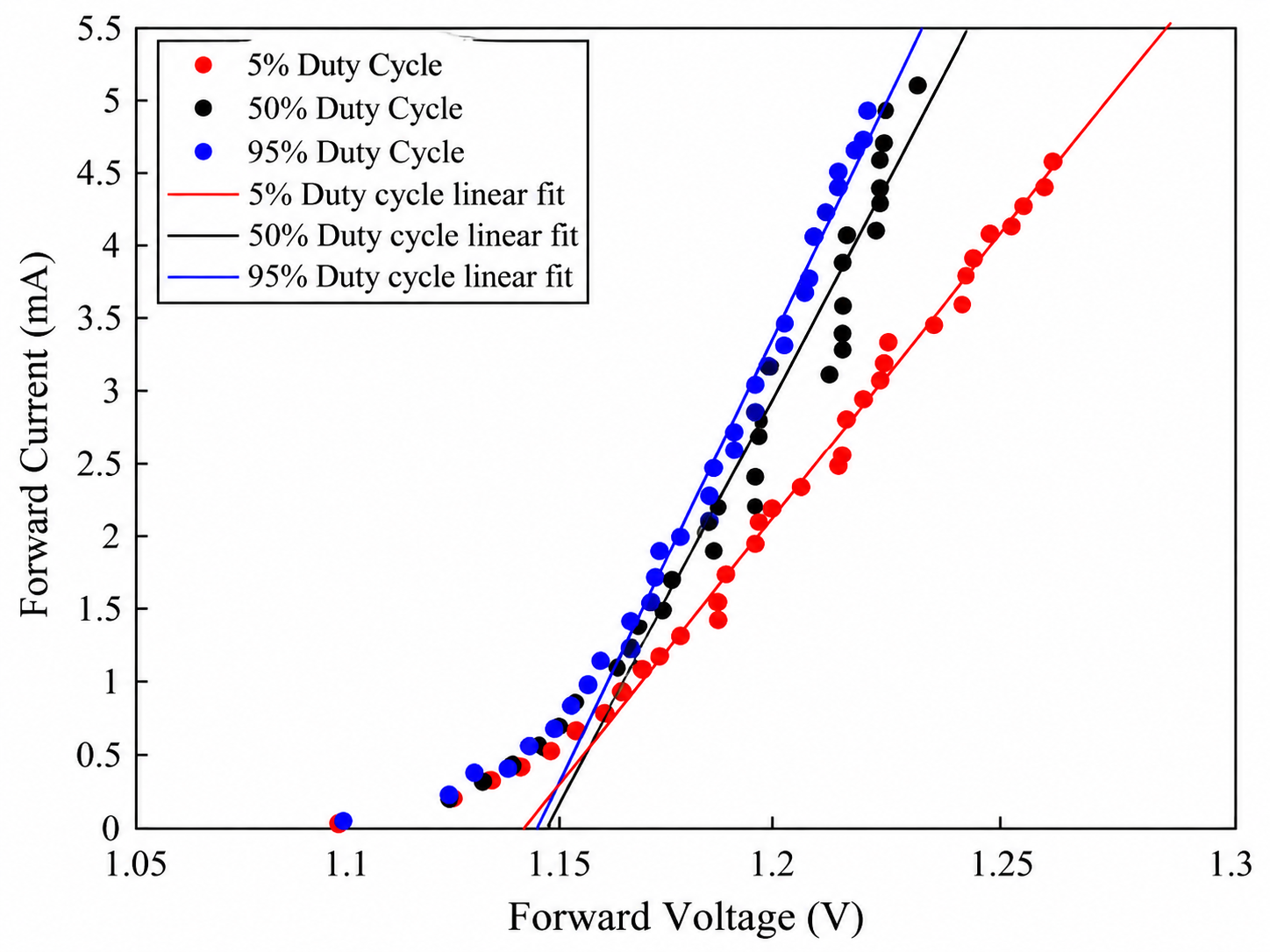}
    \caption{Comparison of forward IV characteristics at a fixed bath temperature measured with different pulse duty cycles (5\%, 50\%, and 95\%), illustrating the influence of self-heating on the extracted slope (apparent resistance).}
\label{fig:dutycycle_iv}
\end{figure}

To evaluate the effect of LED self-heating on these results, we measured at 7~K using a pulsed bias with duty cycles from 5\% to 95\%, while keeping the cycle time constant at 1~ms. At low duty cycle, the LED has sufficient off-time to re-thermalize between pulses, so self-heating is minimal and the IV response reflects the bath temperature. At a high duty cycle, heat accumulates during the on-time if the current is significant, effectively raising the junction temperature and thus increasing the measured slope. This yields an artificially reduced apparent resistance. Accordingly, interpret resistance values extracted under high-duty-cycle operation with caution, since the junction temperature may exceed the bath temperature and may even depend on the current; in that case, the slope does not represent an actual effective resistance under any specific steady-state condition. In contrast, the extracted cut-in voltage remains consistent within uncertainty throughout the tested duty cycles. Figure~\ref{fig:dutycycle_iv} shows the IV curves at 7~K for three different duty cycles.

\subsection{Spectral analysis and discussion}

We recorded the emission spectra of both LEDs at each temperature while driving the devices at their maximum continuous operating current of 3.3~mA. As the temperature decreased, the emission peak shifted systematically toward shorter wavelengths, consistent with the increase in band-gap energy at lower temperatures. Although the emission peak remained clearly identifiable at all temperatures, the spectral profile was distinctly non-Gaussian, exhibiting noticeable asymmetry and a pronounced tail toward shorter wavelengths. This asymmetry may arise from band-filling effects at finite injection levels, inhomogeneous broadening associated with compositional fluctuations in the active region, and contributions from multiple radiative recombination channels \cite{Weisbuch2020}. A single-Gaussian model could not adequately reproduce both the central emission peak and the asymmetric short-wavelength tail, resulting in systematic residuals and potentially biased estimates of the peak position and spectral width. We therefore employed a minimal two-component phenomenological model in which the dominant Gaussian describes the principal emission band, while the subdominant Gaussian accounts for the weaker spectral shoulder or tail.

In most cases, the integrated-area fraction of the subdominant Gaussian was less than 20\% of the total fitted emission, confirming that the principal Gaussian accounts for the majority of the spectral intensity while the secondary component provides a correction for the asymmetric tail. This approach provides a compact, consistent representation of the measured line shape and enables extraction of the peak wavelength, spectral width, and relative integrated contribution of each component. The use of two Gaussians is not intended to establish the presence of exactly two discrete microscopic transitions, but rather to provide a robust empirical description of the asymmetric spectra over the full temperature range.

\begin{figure}[t]
	\centering
	\includegraphics[width=\columnwidth]{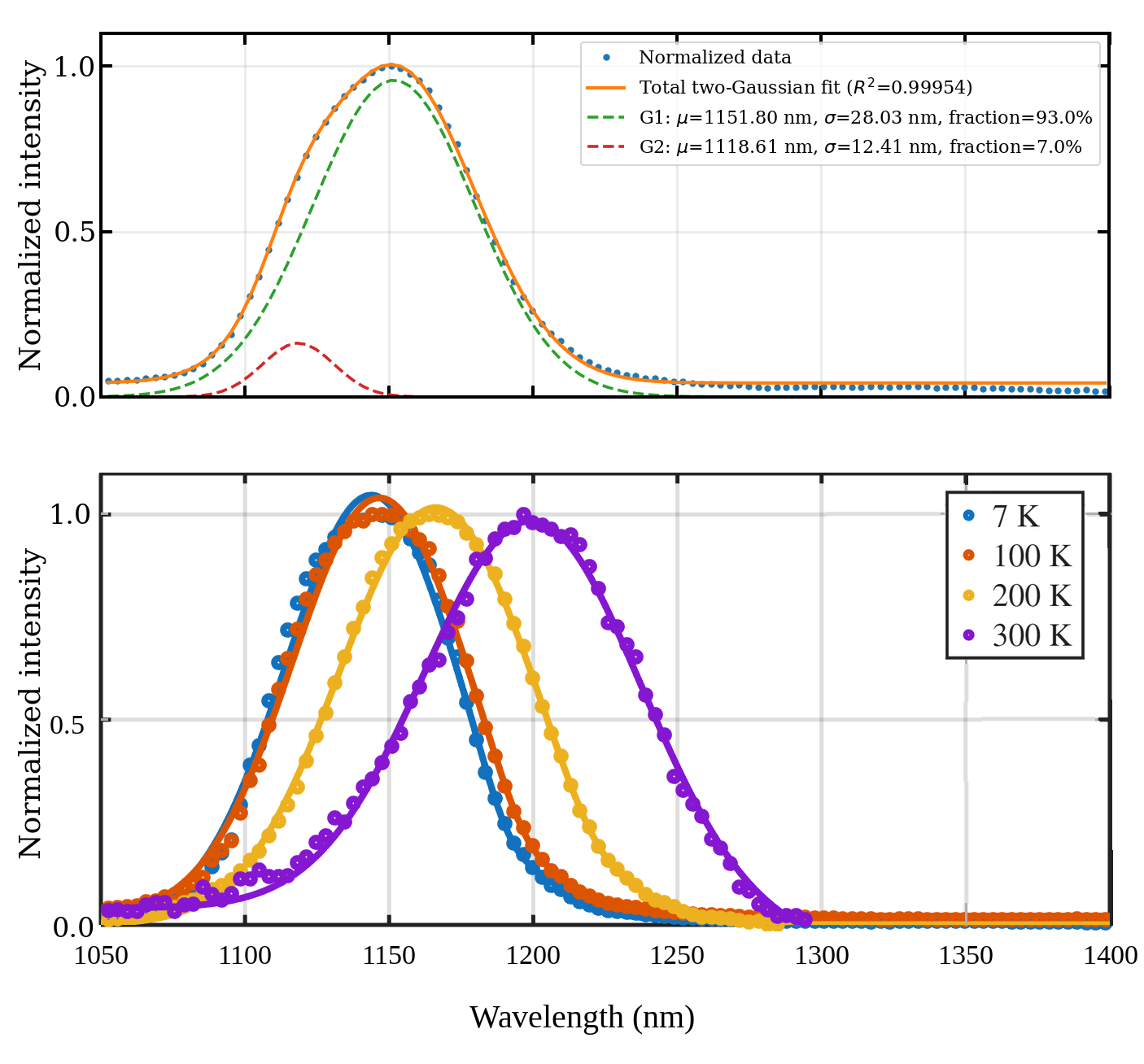}
    \vspace{0.2cm} 

 \caption{Temperature-dependent emission spectra of the 1200~nm LED. The top panel shows the normalized spectrum measured at 20~K together with a two-Gaussian fit. The total fit ($R^2=0.99954$) is decomposed into a dominant component centered at $\mu=1151.80$~nm with $\sigma=28.03$~nm and a secondary component centered at $\mu=1118.61$~nm with $\sigma=12.41$~nm, contributing 93.0\% and 7.0\% of the integrated Gaussian area, respectively. The bottom panel shows normalized emission spectra measured at 7, 100, 200, and 300~K, illustrating the systematic peak shift with temperature.}
 
 \label{fig:spectra_examples}
\end{figure}

The reported peak position corresponds to the mean of the primary (dominant) Gaussian component, which provides a consistent measure of the band-edge emission. Representative normalized spectra of the 1200-nm LED at four selected temperatures are shown in Fig. \ref{fig:spectra_examples}.

\begin{figure}[t]
	\centering
	\includegraphics[width=\columnwidth]{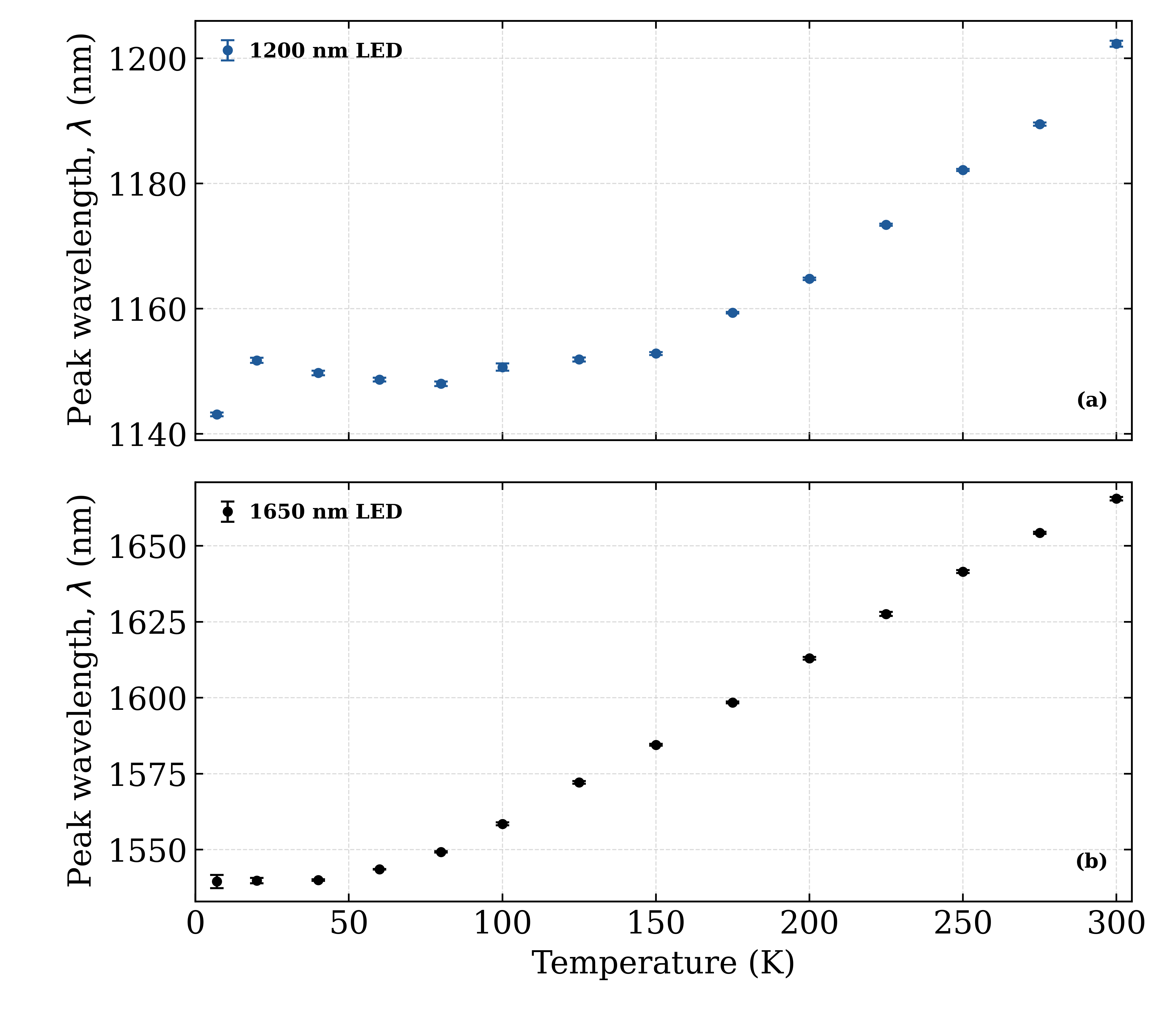}
    \caption{Temperature dependence of the extracted peak wavelength for the nominal 1200~nm LED (top) and 1650~nm LED (bottom). The vertical error bars represent the uncertainties in the fitted peak positions; in most cases, they are smaller than the marker size and are therefore not visible. Here, the uncertainties are on the mean, whereas in Fig.~4 of Ref.~\cite{Ghaith2020} the standard deviation was used incorrectly.}
	\label{fig:lambda_vs_T}
\end{figure}

The extracted peak wavelength at each operating temperature is shown in Fig. \ref{fig:lambda_vs_T}. The error bars represent the standard deviation of the dominant Gaussian. At 300 K, the peak wavelength was measured as 1202~nm (1665~nm) and shifted to 1143~nm (1540~nm) at 7~K for the 1200-nm (1650-nm) LED, corresponding to a total shift of 59~nm (125~nm), or about 5\% (8\%) (see Tab.~\ref{tab:lambda1} and Tab.~\ref{tab:lambda2})

\begin{table*}[t]

\caption{Parameters of the 1200-nm LED extracted from two-Gaussian fits to the emission spectra.
$\lambda_1$ and $\sigma_1$ denote the peak wavelength and standard deviation of the dominant Gaussian component, respectively, while $\lambda_2$ and $\sigma_2$ correspond to the second Gaussian component.
$F$ is the fractional contribution of the dominant Gaussian to the total emission spectrum.}

\label{tab:lambda1}

\centering
\small
\setlength{\tabcolsep}{8pt}
\renewcommand{\arraystretch}{1.15}

\begin{tabular}{cccccc}
\hline
\textbf{T (K)} &
\textbf{$\lambda_1$ (nm)} &
\textbf{$\sigma_1$ (nm)} &
\textbf{$\lambda_2$ (nm)} &
\textbf{$\sigma_2$ (nm)} &
\textbf{$F$ (\%)} \\
\hline

7   & 1143.1(0.3) & 28.9(0.4) & 1247.0(13.2) & 22.1(15.7) & 98 \\
20  & 1151.8(0.4) & 28.0(0.3) & 1118.6(0.8) & 12.4(0.8) & 93 \\
40  & 1149.7(0.4) & 28.8(0.3) & 1116.8(0.6) & 11.7(1.0) & 95 \\
60  & 1148.7(0.3) & 28.6(0.2) & 1116.0(0.4) & 12.7(0.7) & 94 \\
80  & 1148.0(0.4) & 29.1(0.2) & 1116.9(0.6) & 11.2(1.0) & 95 \\
100 & 1150.7(0.6) & 23.4(0.7) & 1122.6(1.5) & 18.7(0.6) & 68 \\
125 & 1151.9(0.3) & 29.5(0.2) & 1122.3(0.8) & 10.2(1.2) & 96 \\
150 & 1152.8(0.2) & 30.6(0.2) & 2242.0(18.0) & 91.0(19.0) & 98 \\
175 & 1159.4(0.1) & 32.5(0.1) & 2196.0(13.6) & 95.2(14.8) & 95 \\
200 & 1164.8(0.2) & 32.3(0.1) & 1192.9(1.1) & 10.1(1.5) & 98 \\
225 & 1173.4(0.2) & 32.1(0.5) & 1169.8(3.4) & 70.7(10.0) & 78 \\
250 & 1182.1(0.2) & 32.8(0.3) & 1153.1(12.3) & 124.7(23.5) & 54 \\
275 & 1189.5(0.3) & 33.9(0.5) & 1206.1(8.6) & 118.8(24.6) & 54 \\
300 & 1202.3(0.5) & 33.9(0.8) & 1179.6(4.5) & 67.1(4.6) & 64 \\

\hline
\end{tabular}
\end{table*}
\begin{table*}[t]
\caption{Parameters of the 1650-nm LED extracted from two-Gaussian fits to the emission spectra. 
$\lambda_1$ and $\sigma_1$ denote the peak wavelength and standard deviation of the dominant Gaussian component, respectively, while $\lambda_2$ and $\sigma_2$ correspond to the second Gaussian component. 
$F$ is the fractional contribution of the dominant Gaussian to the total emission spectrum.}
\label{tab:lambda2}

\centering
\small
\setlength{\tabcolsep}{8pt}   
\renewcommand{\arraystretch}{1.15}

\begin{tabular}{cccccc}
\hline
\textbf{T (K)} &
\textbf{$\lambda_1$ (nm)} &
\textbf{$\sigma_1$ (nm)} &
\textbf{$\lambda_2$ (nm)} &
\textbf{$\sigma_2$ (nm)} &
\textbf{$F$ (\%)} \\
\hline
7   & 1539.5(2.2) & 32.4(0.7) & 1509.3(1.7) & 22.9(2.6) & 89 \\
20  & 1539.8(0.9) & 31.8(0.3) & 1508.9(0.9) & 20.4(1.7) & 93 \\
40  & 1540.0(0.3) & 32.9(0.2) & 1509.5(0.9) & 17.3(1.7) & 97 \\
60  & 1543.6(0.1) & 34.0(0.1) & 1514.3(1.3) & 11.9(1.8) & 99 \\
80  & 1549.3(0.2) & 34.9(0.4) & 1538.9(4.9) & 63.9(16.6) & 89 \\
100 & 1558.5(0.5) & 34.6(0.6) & 1546.5(3.2) & 50.9(4.6) & 75 \\
125 & 1572.1(0.4) & 35.1(0.5) & 1554.2(2.4) & 52.5(2.5) & 69 \\
150 & 1584.5(0.3) & 35.3(0.3) & 1560.9(1.7) & 53.1(1.2) & 63 \\
175 & 1598.5(0.3) & 36.4(0.4) & 1565.3(3.2) & 53.3(1.3) & 65 \\
200 & 1613.0(0.4) & 37.0(0.8) & 1568.0(11.8) & 51.8(3.9) & 67 \\
225 & 1627.6(0.6) & 36.9(1.2) & 1579.9(15.1) & 52.1(5.2) & 63 \\
250 & 1641.5(0.5) & 37.0(1.3) & 1593.2(13.7) & 54.1(4.8) & 59 \\
275 & 1654.2(0.3) & 37.8(1.6) & 1606.4(16.1) & 57.3(5.7) & 56 \\
300 & 1665.5(0.6) & 40.0(2.4) & 1618.9(25.3) & 62.4(9.0) & 54 \\
\hline
\end{tabular}
\end{table*}
The peak shifts down linearly between 300~K and approximately 80~K, at which point the temperature dependence weakens. This behavior is consistent with Varshni's description of the band-gap temperature dependence (Eq.(\ref{eq:Varnish})).
For temperatures $T\ll\beta$, the dependence becomes approximately quadratic in $T$, and the slope $dE_g/dT$ approaches zero as $T$ goes to zero. The observed saturation of the peak wavelength below 80~K therefore reflects the intrinsic freezing out of electron–phonon interactions and the approach of the band gap to its low-temperature limit \cite{Villegas2017}. The measured redshift with increasing temperature is qualitatively consistent with Varshni-type band-gap narrowing in InGaAsP. Near room temperature, the measured wavelength coefficients are comparable to values reported for long-wavelength InGaAsP/InP emitters, although the full 7–300 K behavior is not quantitatively described by a single set of bulk Varshni parameters \cite{Satzke1988, varshni1967_energygap}.

\section{Conclusion}
\label{sec:conclusion}

The low-temperature IV characteristics of the InGaAsP LED under investigation exhibit the expected diode behavior, with the cut-in voltage increasing substantially as the temperature decreases from 300~K to 7~K, qualitatively consistent with the expected band-gap widening \cite{SHEN2019106179, temp_dep}. In contrast, the apparent series resistance shows an anomalous temperature dependence at low temperatures, likely influenced by self-heating under continuous bias. Pulsed IV measurements support this interpretation, as duty-cycle variation alters the low-temperature curve shape while leaving the extracted cut-in voltage largely unchanged.

Optical measurements reveal a systematic blue shift of approximately 5\% (8\%) between 300~K and 7~K for the 1200~nm (1650~nm) LED. The extracted emission energies follow Varshni-type behavior \cite{varshni1967_energygap} and are in good agreement with the electrical analysis, providing a consistent description of the temperature-dependent evolution of the band gap.
\clearpage
\bibliography{apssamp}

@book{kittel2018introduction,
  title         = {Introduction to Solid State Physics},
  author        = {Kittel, Charles and McEuen, Paul},
  year          = {2018},
  publisher     = {John Wiley \& Sons}
}

@article{odonnell1991_bandgaps,
  title         = {Temperature dependence of semiconductor band gaps},
  author        = {O’donnell, Kevin P. and Chen, Xie},
  journal       = {Applied Physics Letters},
  volume        = {58},
  number        = {25},
  pages         = {2924--2926},
  year          = {1991},
  publisher     = {American Institute of Physics}
}

@article{varshni1967_energygap,
 title         = {Temperature dependence of the energy gap in semiconductors},
  author        = {Varshni, Yatendra Pal},
  journal       = {Physica},
  volume        = {34},
  number        = {1},
  pages         = {149--154},
  year          = {1967},
  publisher     = {Elsevier}
}

@article{reifegerste2008_ledspectra,
   title         = {Modelling of the temperature and current dependence of {LED} spectra},
  author        = {Reifegerste, Frank and Lienig, Jens},
  journal       = {Journal of Light \& Visual Environment},
  volume        = {32},
  number        = {3},
  pages         = {288--294},
  year          = {2008},
  publisher     = {The Illuminating Engineering Institute of Japan}
}

@article{wu2012_ingan_el,
   title         = {Thermal effect on the electroluminescence of {InGaN/GaN} multiquantum-well light-emitting devices},
  author        = {Wu, Ya-Fen and Hsu, Hung-Pin and Liu, Tsung-Yu},
  journal       = {Solid-State Electronics},
  volume        = {68},
  pages         = {63--67},
  year          = {2012},
  publisher     = {Elsevier}
}

@article{fulmek2017_inganfwd,
title         = {Time dependent and temperature dependent properties of the forward voltage characteristic of {InGaN} high power {LEDs}},
  author        = {Fulmek, P. L. and Haumer, P. and Wenzl, F. P. and Nemitz, W. and Nicolics, J.},
  journal       = {AIP Advances},
  volume        = {7},
  number        = {3},
  year          = {2017},
  publisher     = {AIP Publishing}
}

@article{akimoto1998_ledlowT,
   title         = {Forward current--voltage characteristics of light-emitting diodes at low temperatures},
  author        = {Akimoto, H. and Marchenkov, A. and Jochemsen, R. and Frossati, G.},
  journal       = {Cryogenics},
  volume        = {38},
  number        = {4},
  pages         = {451--452},
  year          = {1998},
  publisher     = {Elsevier}
}

@book{neamen2012_semiconductor,
   author        = {Neamen, Donald A.},
  title         = {Semiconductor Physics and Devices: Basic Principles},
  edition       = {4},
  publisher     = {McGraw-Hill Education},
  year          = {2012},
  isbn          = {9780073529585},
  note          = {See pp.~241--253.}
}

@misc{SMC1200datasheet,
  author       = {{Marubeni Co.}},
  title        = {{SMC1200 InGaAsP LED} Datasheet},
  howpublished = {\url{https://www.tech-led.com/wp-content/uploads/2015/05/SMC1200.pdf}},
  year         = {n.d.},
  note         = {Undated datasheet; accessed October 2025}
}

@article{LI2000215,
title = {Material parameters of InGaAsP and InAlGaAs systems for use in quantum well structures at low and room temperatures},
journal = {Physica E: Low-dimensional Systems and Nanostructures},
volume = {5},
number = {4},
pages = {215-273},
year = {2000},
issn = {1386-9477},
doi = {https://doi.org/10.1016/S1386-9477(99)00262-3},
url = {https://www.sciencedirect.com/science/article/pii/S1386947799002623},
author = {E.Herbert Li},
}

@article{5lnp-6mng,
  title = {Search for low-mass electron-recoil dark matter using a single-charge sensitive SuperCDMS-HVeV detector},
  author = {Albakry \textit{et al.} (SuperCDMS Collaboration), M. F.},
  collaboration = {SuperCDMS Collaboration},
  journal = {Phys. Rev. D},
  volume = {113},
  issue = {3},
  pages = {032001},
  numpages = {10},
  year = {2026},
  month = {Feb},
  publisher = {American Physical Society},
  doi = {10.1103/5lnp-6mng},
  url = {https://link.aps.org/doi/10.1103/5lnp-6mng}
}

@article{PhysRevD.111.012006,
  title = {Light dark matter constraints from SuperCDMS HVeV detectors operated underground with an anticoincidence event selection},
  author = {Albakry \textit{et al.} (SuperCDMS Collaboration), M. F.},
  journal = {Phys. Rev. D},
  volume = {111},
  issue = {1},
  pages = {012006},
  numpages = {9},
  year = {2025},
  month = {Jan},
  publisher = {American Physical Society},
  doi = {10.1103/PhysRevD.111.012006},
  url = {https://link.aps.org/doi/10.1103/PhysRevD.111.012006}
}

@article{Kennard:2026xlx,
    author = "Kennard, Kyle and others",
    title = "{Performance of a SuperCDMS HVeV detector with Sub-eV energy resolution and single charge-sensitivity}",
    eprint = "2601.16307",
    archivePrefix = "arXiv",
    primaryClass = "physics.ins-det",
    doi = "10.1016/j.nima.2026.171753",
    journal = "Nucl. Instrum. Meth. A",
    volume = "1091",
    pages = "171753",
    year = "2026"
}

@article{band,
    author = {Vurgaftman, I. and Meyer, J. R. and Ram-Mohan, L. R.},
    title = {Band parameters for III–V compound semiconductors and their alloys},
    journal = {Journal of Applied Physics},
    volume = {89},
    number = {11},
    pages = {5815-5875},
    year = {2001},
    month = {06},
    issn = {0021-8979},
    doi = {10.1063/1.1368156},
    url = {https://doi.org/10.1063/1.1368156},
}

@article{band2,
title = {A study of energy band gap versus temperature for Cu2ZnSnS4 thin films},
journal = {Physica B: Condensed Matter},
volume = {407},
number = {1},
pages = {108-111},
year = {2012},
issn = {0921-4526},
doi = {https://doi.org/10.1016/j.physb.2011.09.134},
url = {https://www.sciencedirect.com/science/article/pii/S0921452611010374},
author = {Prashant K. Sarswat and Michael L. Free},
}

@article{PhysRevB.95.035211,
  title = {Low-temperature semiconductor band-gap thermal shifts: ${T}^{4}$ shifts from ordinary acoustic and ${T}^{2}$ from piezoacoustic coupling},
  author = {Allen, Philip B. and Nery, Jean Paul},
  journal = {Phys. Rev. B},
  volume = {95},
  issue = {3},
  pages = {035211},
  numpages = {9},
  year = {2017},
  month = {Jan},
  publisher = {American Physical Society},
  doi = {10.1103/PhysRevB.95.035211},
  url = {https://link.aps.org/doi/10.1103/PhysRevB.95.035211},
}

@article{Ryu:20,
author = {Han-Youl Ryu and Geun-Hwan Ryu and Chibuzo Onwukaeme and Byongjin Ma},
journal = {Opt. Express},
number = {19},
pages = {27459--27472},
publisher = {Optica Publishing Group},
title = {Temperature dependence of the Auger recombination coefficient in InGaN/GaN multiple-quantum-well light-emitting diodes},
volume = {28},
month = {Sep},
year = {2020},
url = {https://opg.optica.org/oe/abstract.cfm?URI=oe-28-19-27459},
doi = {10.1364/OE.402831},
}

@article{Rashidi2019,
   author = {Arman Rashidi and Morteza Monavarian and Andrew Aragon and Daniel Feezell},
   doi = {10.1038/s41598-019-56390-2},
   issn = {20452322},
   issue = {1},
   journal = {Scientific Reports},
   month = {12},
   pmid = {31882667},
   publisher = {Nature Research},
   title = {Thermal and efficiency droop in InGaN/GaN light-emitting diodes: decoupling multiphysics effects using temperature-dependent RF measurements},
   volume = {9},
   year = {2019}
}

@misc{Weisbuch2020,
   author = {Claude Weisbuch and Shuji Nakamura and Yuh Renn Wu and James S. Speck},
   doi = {10.1515/nanoph-2020-0590},
   issn = {21928614},
   issue = {1},
   journal = {Nanophotonics},
   month = {1},
   pages = {3-21},
   publisher = {De Gruyter Open Ltd},
   title = {Disorder effects in nitride semiconductors: Impact on fundamental and device properties},
   volume = {10},
   year = {2020}
}

@article{Villegas2017,
  author  = {Villegas, Cesar E. P. and Rocha, A. R. and Marini, Andrea},
  title   = {Anomalous Temperature Dependence of the Band Gap in Black Phosphorus},
  journal = {Nano Letters},
  volume  = {16},
  number  = {8},
  pages   = {5095--5101},
  year    = {2016},
  doi     = {10.1021/acs.nanolett.6b02035}
}

@article{SHEN2019106179,
title = {The investigation of temperature dependent electrical characteristics of {Au/Ni/$\beta$-(InGa)$_2$O$_3$} Schottky diode},
journal = {Superlattices and Microstructures},
volume = {133},
pages = {106179},
year = {2019},
issn = {0749-6036},
doi = {https://doi.org/10.1016/j.spmi.2019.106179},
url = {https://www.sciencedirect.com/science/article/pii/S0749603619304975},
author = {Yixian Shen and Qian Feng and Ke Zhang and Zhuangzhuang Hu and Guangshuo Yan and Yuncong Cai and Wenxiang Mu and Zhitai Jia and Chunfu Zhang and Hong Zhou and Jincheng Zhang and Xiaozheng Lian and Zhanping Lai and Yue Hao},
}

@article{temp_dep,
  author  = {Nippert, Felix and Karpov, Sergey Yu. and Callsen, Gordon and Galler, Bastian and Kure, Thomas and Nenstiel, Christian and Wagner, Markus R. and Stra{\ss}burg, Martin and Lugauer, Hans-J{\"u}rgen and Hoffmann, Axel},
  title   = {Temperature-dependent recombination coefficients in {InGaN} light-emitting diodes: Hole localization, {Auger} processes, and the green gap},
  journal = {Applied Physics Letters},
  volume  = {109},
  number  = {16},
  pages   = {161103},
  year    = {2016},
  doi     = {10.1063/1.4965298}
}

@article{Satzke1988,
   author = {K Satzke and G Weiser and R Höger and W Thulke},
   doi = {10.1063/1.340373},
   issn = {0021-8979},
   issue = {11},
   journal = {Journal of Applied Physics},
   month = {6},
   pages = {5485-5490},
   title = {Absorption and electroabsorption spectra of an {In$_{1-x}$Ga$_{x}$P$_{1-y}$As$_{y}$/InP} double heterostructure},
   volume = {63},
   url = {https://doi.org/10.1063/1.340373},
   year = {1988}
}

@article{DelCastello2024,
  author  = {Del Castello, Giorgio},
  title   = {{LANTERN}: A Multichannel Light Calibration System for
             Cryogenic Detectors},
  journal = {Nuclear Instruments and Methods in Physics Research Section A},
  volume  = {1068},
  pages   = {169728},
  year    = {2024},
  doi     = {10.1016/j.nima.2024.169728}
}

@article{Ghaith2020,
  author  = {Ghaith, M. and Rau, W. and Peterson-Galema, M. and
             Di Stefano, P. and Fascione, E. and Germond, R. and
             Underwood, R.},
  title   = {New Approaches to Low-Energy Calibration of Cryogenic Detectors},
  journal = {Journal of Low Temperature Physics},
  volume  = {200},
  pages   = {305--311},
  year    = {2020},
  doi     = {10.1007/s10909-020-02469-4}
}

@article{XENON2026LightDM,
  author    = {{XENON Collaboration}},
  title     = {Light Dark Matter Search with 7.8 Tonne-Year of Ionization-Only Data in XENONnT},
  journal   = {Physical Review Letters},
  volume    = {137},
  number    = {5},
  pages     = {051003},
  year      = {2026},
  doi       = {10.1103/2lrq-f6bk},
  publisher = {American Physical Society}
}

\end{document}